\documentclass[preprint]{aastex}

\newcommand{\photu}{ph\ cm$^{-2}$ s$^{-1}$ sr$^{-1}$ \AA$^{-1}$}
\newcommand{\galex}{{\it GALEX}}

\usepackage{hyperref}
\usepackage{amsmath}
\usepackage{dirtytalk}
\usepackage{csquotes}

\shorttitle{MULTI-WAVELENGTH CORRELATION STUDIES}
\shortauthors{Jyothy et al.}

\begin{document}

\title{MULTI-WAVELENGTH CORRELATION STUDIES OF DIFFUSE EMISSION FROM THE AQUILA RIFT}

\author{S. N. Jyothy\altaffilmark{1}, N. V. Sujatha\altaffilmark{2}, and Narayanankutty Karuppath\altaffilmark{1}}

\altaffiltext{1}{Department of Physics, Amrita Vishwa Vidyapeetham, Amritapuri, India}
\altaffiltext{2}{St. Xavier's College for Women, Aluva - 683101, India.}

\begin{abstract}

We present a multi-wavelength correlation study of diffuse ultraviolet radiation using {\it GALEX} observations towards the Aquila Rift. Apart from airglow and zodiacal emissions, we find a diffuse background of 1300 - 3700 \photu\ in the far-ultraviolet (FUV: 1350 - 1750 \AA) band and 1300 - 2800 \photu\ in the near-ultraviolet (NUV: 1750 - 2850 \AA) band. The observed diffuse UV emissions are saturated with total as well as neutral hydrogen column density in the region due to high optical depth in UV ($\tau$: 0.91-23.38). Higher values of FUV/NUV ratio in the region, greater than the threshold value of 0.6, along with the positive correlation between the ratio and FUV intensity are due to excess emission in the FUV band which is absent in the NUV band. We estimated the excess emission to be in the range $\sim$ 400 - 2700 \photu, plausibly due to H$_2$ fluorescence, ion line emissions and two-photon continuum emissions from the region in the FUV band, which also shows saturation in optically thick regions with N(H$_{2}$) as well as H$\alpha$ emissions. Since N(H$_{2}$) and H$\alpha$ emissions spread all over the region, the excess emission from the field is composite in nature and a detailed spectroscopic analysis is needed to disentangle the contribution from individual components.

\end{abstract}


\section{Introduction}

The diffuse ultraviolet (UV) radiation is a complex mixture of astrophysical signal \citep{hay69, lil76} mixed with zodiacal light and airglow \citep{murthy14b} with some contribution from extragalactic background at high galactic latitudes \citep{suj10}. But, the dust scattered starlight forms majority of the diffuse UV radiation in the low to mid-galactic latitudes. Early UV observations mainly focused on high galactic latitudes for the purpose of studying the extragalactic background in the sky \citep{sas95, schim01}. This limited the studies of diffuse UV observations towards low latitude regions. Spectroscopy of Plasma Evolution from Astrophysical Radiation ({\it SPEAR}; \citet{edel06}) is one among the few missions which conducted spectral imaging survey in UV at wavelength range of 900 - 1750 \AA\  towards low latitudes. Another such observation came from the Galaxy Evolution Explorer ({\it GALEX}; \citet{martin05}; \citet{morr07}), which had very few observations at low latitudes taken during the end stage of the mission by 2013. Majority of them were Near Ultraviolet (NUV) observations since the Far Ultraviolet (FUV) detector power supply had failed by then.

Using {\it SPEAR} and {\it GALEX} data, \citet{park12} analysed the diffuse UV background from the optically thin regions of Aquila Rift with visual extinction A$_v$ $\le$ 1. They reported that dust-scattered starlight forms majority of the diffuse FUV emission and also indicated the presence of molecular hydrogen fluorescence as well as the ion line emissions and two-photon continuum emissions. \citet{jyo15} studied the diffuse UV radiation from the Aquila Rift using {\it GALEX} data and modeled the observed emission in FUV and NUV bands. Here we present the correlation analysis of diffuse UV background from the Aquila region with multi-wavelength emissions such as IR 100 $\micron$, N(H), H$\alpha$, N(\ion{H}{1}) and N(CO) for a better understanding of the region.

Aquila Rift is a dark molecular cloud covering parts of constellations of Aquila, Serpens Cauda and Eastern Ophiuchus; the structure of which is clearly visible in H$\alpha$ images \citep{madsen05}. The Galactic CO surveys reveal that the Aquila Rift is irregular in shape, spanning 227 square degrees ({\it l} : 15{\degr} to 40{\degr} and  {\it b}: -6{\degr} to +20{\degr}) in the sky. The distance to the Aquila Rift as estimated by \citet{str03} is 225$\pm$25 pc from the sun with a thickness of about 80 pc. They also set the limit of the extinction close to 3.0 mag towards the region. The optical depth of the region in FUV is estimated as 1.0 to 23 using the relation: $\tau (1539\ \text{\AA})$ = $N(H) \times \sigma (1539\ \text{\AA})$, where $\sigma$ is the dust extinction cross-section. The interstellar radiation fields are not uniform \citep{park12} in this region, towards which a significant depression of diffuse UV emission has already been reported.

\section{Observations and Data}

{\it GALEX}, a low earth orbiting UV telescope was launched in 2003 under NASA's Small Explorer (SMEX) program. The spacecraft used a single, 50 cm Ritchey - Chr\'{e}tien telescope to collect light from the sky in two wavelength bands - FUV (1516 \AA) and NUV (2267 \AA) - with a spatial resolution of $5\arcsec$ - $10\arcsec$. The main goal of the mission was to observe galaxies at low redshifts. The instrument conducted all sky imaging surveys with a $1.25\degr$ circular field of view. From June 7, 2003 to June 2, 2013, {\it GALEX} collected about 44,843 high quality observations which form the final and latest data release, the GR6/GR7. During the initial stage of the mission, low galactic latitudes and brighter regions of the sky were avoided to prevent damage of the detectors. But towards the end of the mission, the spacecraft observed the above said regions although the FUV detector had stopped working by then. Thus the final data release has a wealth of observations at low galactic latitudes and brighter regions, but the number of FUV observations is less when compared to that of NUV observations. The data set of {\it GALEX} is made publicly available from the Mikulski Archive for Space Telescopes (MAST)\footnote{http://{\it GALEX}.stsci.edu/GR6/}.

Each {\it GALEX} observation comprises of one or more visits which may have been taken on different days, months or even years towards a single location in the sky with differing contributions from zodiacal light and airglow \citep{murthy14a} during each visit. In order to extract the diffuse background, the contributions from point sources and the foreground emissions (airglow and zodiacal light) should be eliminated. \citet{murthy14b} did this for the GR6/GR7 {\it GALEX} data. The point sources were blanked out from each observation using the corresponding merged catalog and the pixels were binned into 2\arcmin\ pixels. Using the empirical formulae derived by \citet{murthy14a}, the airglow and the zodiacal light components of the foreground contribution were calculated and subtracted from each binned pixels.  Thus the final data (refer \citet{murthy14b} for a full description) consist of the astrophysical signal alone. The processed data is available at {\url
{https://archive.stsci.edu/prepds/uv-bkgd/}}. This background is free from edge effects since the central data (0.5\degr\ radius) alone is taken.

Our region constitutes a 15 deg$^2$ area centered at {\it l} = $30\degr$, {\it b} = $10\degr$. There are a total of 227 FUV and 258 NUV observations in the region, the majority being All Sky Imaging Surveys (AIS - refer \citet{jyo15} for a map of {\it GALEX} coverage in the region). We had re-binned the {\it GALEX}\ FUV and NUV data into 6\arcmin\ pixel bins to match the resolutions of other telescopic data and to increase the signal to noise level. This was done by weighting each pixel by the corresponding exposure time. We used this re-binned data to study the variation of diffuse UV background of the region along with the infrared (IR 100$\micron$) (IRAS, Infrared Astronomy Satellite, \citep{schlegel98}), H$\alpha$ \citep{Fink03}, N(\ion{H}{1}) \citep{dick90} and CO maps \citep{dame01}.

\section{Results and Discussion}

\begin{figure*}
\centering
\includegraphics[width=4in]{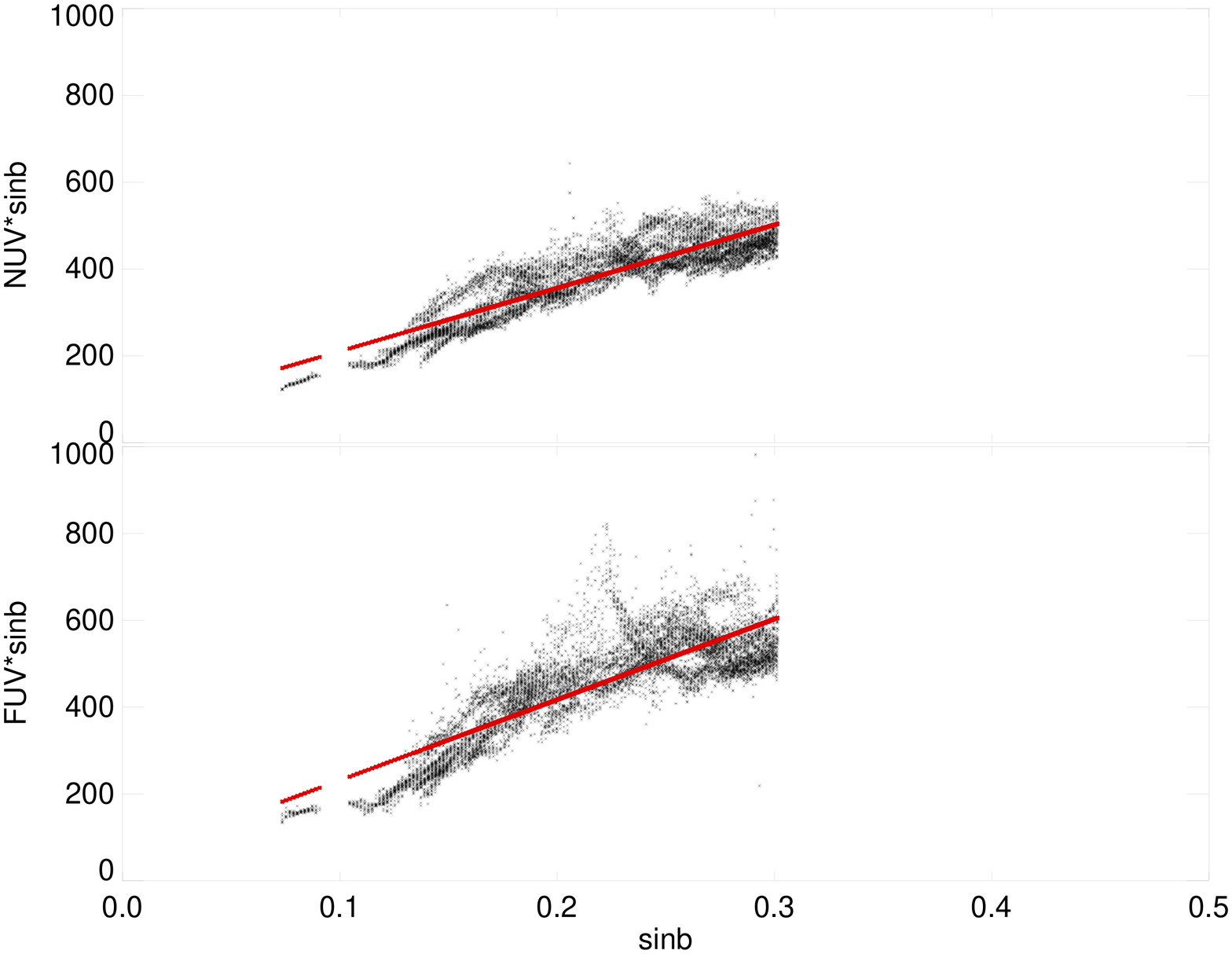}
\caption{Latitude dependence of UV intensities are shown in the figure. The red lines shows the best fit line corresponding to the empirical relations given in Eq.~\ref{eq:1} and Eq.~\ref{eq:2}}
\label{fig1}
\end{figure*}

The observed UV intensities in the region have been corrected for airglow and zodiacal light as
described by \citet{murthy14b}. The estimated FUV and NUV airglow contribution is less than 10 \photu\ (PU: photon units) and zodiacal light in the NUV observations is between 320 - 490 PU \citep{murthy14b}. The median diffuse background in the region is estimated as 2040 PU and 1718 PU with a standard deviation less than 20 PU and 40 PU  respectively in FUV and NUV bands. The latitude dependence of the corrected UV emissions \citep{per91} is shown in Fig.~\ref{fig1} and the empirical relations connecting the UV emissions with galactic latitudes in the region are given by  Eq.~\ref{eq:1} and Eq.~\ref{eq:2}. The observed trend is similar to that reported by \citet{Hamden13}, but with a different slope which is a function of latitude. 

\begin{equation} \label{eq:1}
    FUV\  sin (b)= 1858\  sin (b) + 45
\end{equation}

\begin{equation} \label{eq:2}
    NUV\  sin (b)= 1457\  sin (b) + 64
\end{equation}

\begin{figure}
\centering
\includegraphics[width=4in]{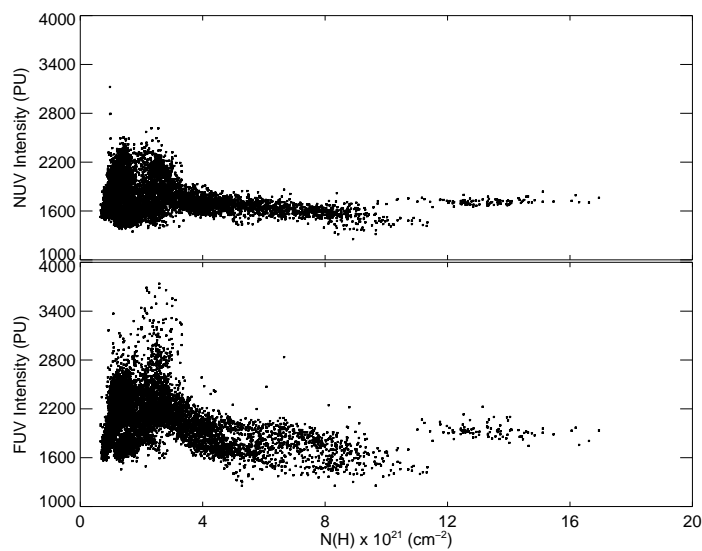}
\caption{{\it GALEX} UV intensities plotted against the total hydrogen column density, N(H). The diffuse background emission shows some correlation in the low dense region and is saturated in optically thick regions}\label{fig2}
\end{figure}

The diffuse UV intensity is mainly due to the scattering of starlight by interstellar dust grains, whereas the IR(100) $\micron$ flux is due to thermal emission from the interstellar dust. \citet{jyo15} found that the diffuse UV emissions in the region is almost saturated with IR 100 $\micron$ emission, as expected for a region of high optical depth in UV, which in turn makes the correlation of UV emissions with the total hydrogen column density (N(H)) implausible. This is because the extinction cross-section of dust grains is much less in IR than in UV implying that the thermal emission observed in the IR band samples a much longer line of sight than that in the UV, which saturates over only a few hundred parsecs. However, N(H) (Fig.~\ref{fig2}) shows a linear correlation with both FUV and NUV in the optically thin regions especially towards \enquote{Ophiuchus} and \enquote{Hercules} where dust extinction is not too high with N(H) $\le$ 2.0 x 10$^{21}\ cm^{-2}$. 

\begin{figure}
\centering
\includegraphics[width=4in]{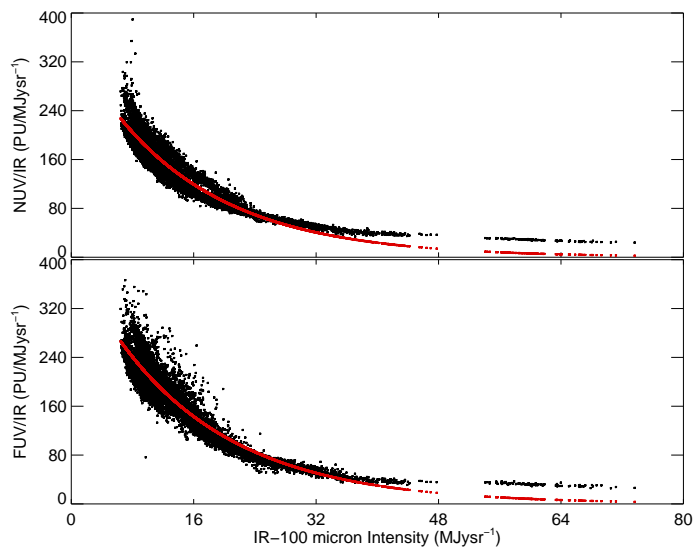}
\caption{Observed UV/IR(100) ratio plotted against IR 100 $\micron$ intensity. The ratio exponentially drops off rapidly with IR(100) due to high optical depth in UV. The red lines indicate the best fit according to the Eq.~\ref{eq:3} and Eq.~\ref{eq:4}}\label{fig3}
\end{figure}

\begin{figure}
\centering
\includegraphics[width=4in]{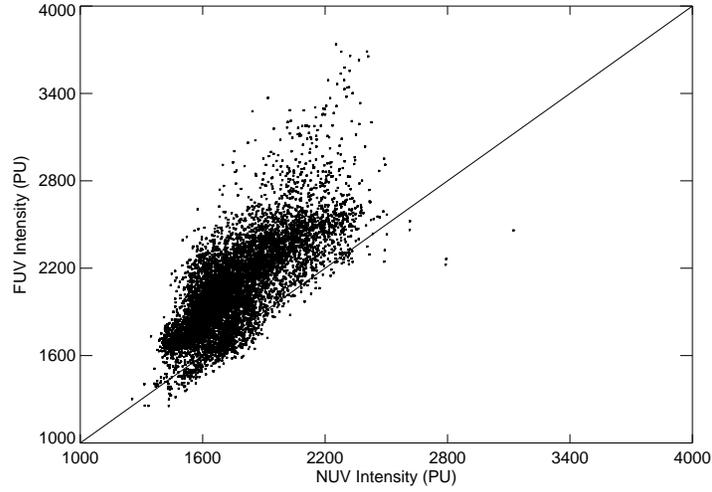}
\caption{FUV emissions plotted against NUV emissions in the region. On comparing the observed intensities with FUV=NUV line (solid line), it becomes clear that FUV intensity does not increase linearly with NUV intensity. This is attributed to excess emissions in the FUV band which is not present in the NUV band.}\label{fig4}
\end{figure}

UV/IR intensity ratio strongly depends on the optical depth and local effects such as the variation of stellar radiation field in the region \citep{suj10}. From Fig.~\ref{fig3}, we see that the UV/FIR ratio drops off exponentially with the IR 100 $\micron$ emission according to Eq.~\ref{eq:3} and Eq.~\ref{eq:4}. It is interesting to note that this curve follows exactly the same curve of \citet{suj10} and \citet{murthy01} even though the UV and IR values are totally different. The strong correlation (r = 0.75) between emissions in the FUV and NUV channels (Fig.~\ref{fig4}) indicates that dust-scattered starlight is the major component of the observed diffuse emission in both bands. However, on comparing the observed intensities with FUV=NUV line (solid line), it becomes clear that FUV intensity does not increase linearly with NUV intensity, with a clear indication of additional emissions in the FUV band (excess FUV emission) which is absent in the NUV band.

\begin{equation} \label{eq:3}
    \frac{FUV}{IR(100)}= 405\ \exp(-0.065\times IR(100))
\end{equation}

\begin{equation} \label{eq:4}
    \frac{NUV}{IR(100)}= 351\ \exp(-0.067\times IR(100))
\end{equation}

\begin{figure}
\centering
\includegraphics[width=4in]{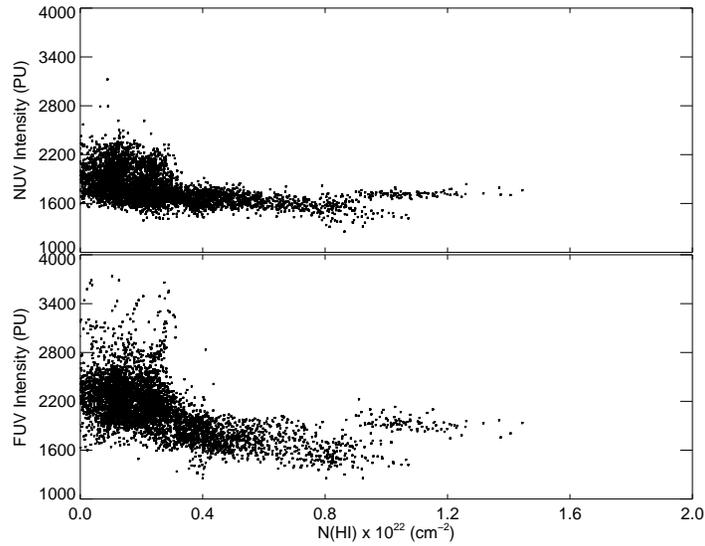}
\caption{UV intensities are plotted against the neutral hydrogen column density, N(\ion{H}{1}). Saturation effect is strong in NUV and is weak in FUV indicating the presence of additional contribution in the FUV emission. }  \label{fig5}
\end{figure}

\begin{figure*}
\centering
\includegraphics[width=4in]{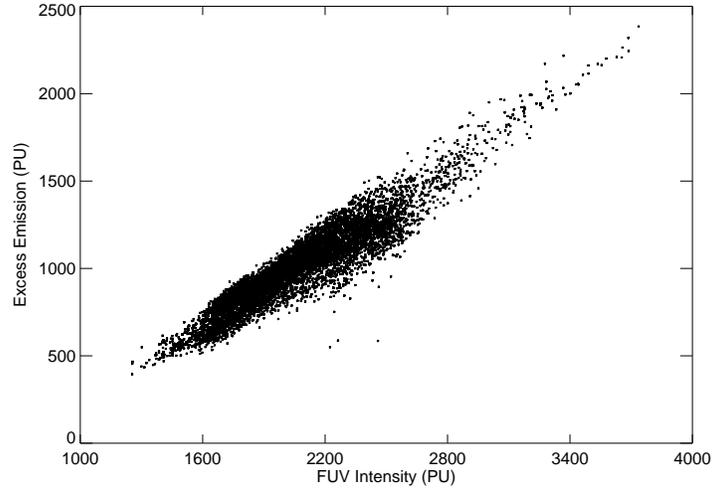}
\caption{Excess FUV emission plotted against the total FUV intensity. Tight correlation indicates the presence of excess emission in the region.}
\label{fig6}
\end{figure*}

\begin{figure}
\centering
\includegraphics[width=4in]{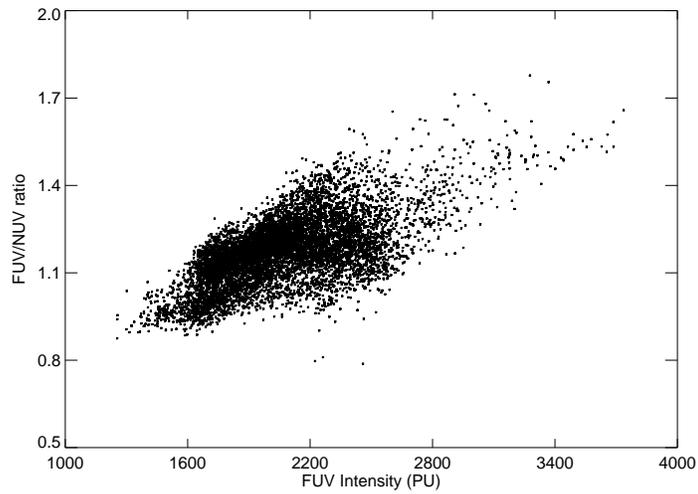}
\caption{FUV/NUV ratio plotted against the FUV intensity. Strong positive correlation along with higher ratio values than the threshold is a clear indication of excess emission in the FUV band}\label{fig7}
\end{figure}

The dust scattered radiation largely depends on the neutral hydrogen column density N(\ion{H}{1}) and to some extent on the local stellar radiation field. However, we could not find any correlation between them (Fig.~\ref{fig5}) due to the high optical depth of the region. \citet{suj10} strongly recommends that the FUV/NUV ratio can be used to identify the atomic and molecular emission regions in the {\galex} survey fields all over the sky and the ratio $\le$ the threshold value, i.e., ratio at $\tau_{\tiny{FUV}}$ \ $\sim$ 0.9, is attributed to the dust-scattered radiation alone and any increase in this ratio is assumed to be due to an \enquote{excess FUV emission} in the region. Following this, the excess emission in FUV is estimated as 400 - 2700 PU over the field by assuming an empirical threshold value of 0.6 for the region. Since the region is optically thick in UV, the dust scattered emissions in the UV bands are saturated. Hence, any change in the total FUV background is the excess emission which shows a tight correlation with the total FUV background (Fig.~\ref{fig6}). High values of UV ratio between two bands over the region along with a reasonably good correlation (r = 0.64) with the FUV (Fig.~\ref{fig7}) indicates the presence of excess emission in FUV band which is absent in NUV band. The possible contributors to the excess FUV emission are H$_{2}$ fluorescence, two photon continuum emissions, and line emission from ionized gas (in the temperature range of $10^4 -10^6$ K) in the interstellar medium \citep{suj10,sha06, park12}. It is also clear from Fig.~\ref{fig2} that, the excess emission is more intense in the relatively low-density regions where the FUV emissions are maximum.

\begin{figure*}
\centering
\includegraphics[width=4in]{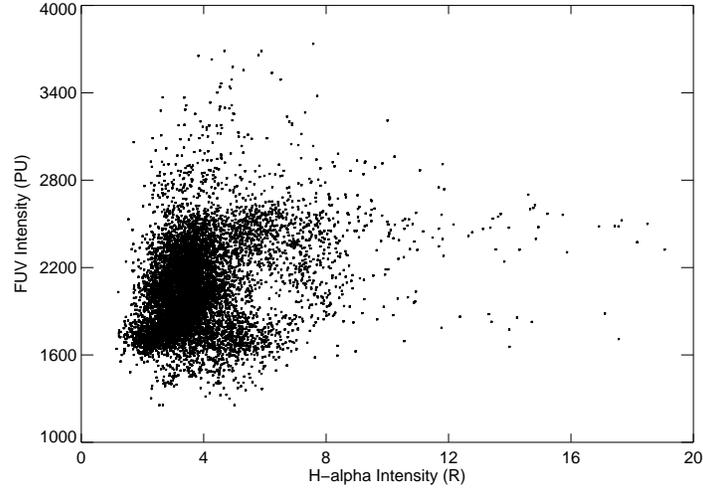}
\caption{Total FUV emission in the region plotted against H$\alpha$ intensity. The FUV emission shows some positive correlation with H$\alpha$ emission especially in the faint H$\alpha$ regions and the correlation decreases as the H$\alpha$ intensity increases.}
\label{fig8}
\end{figure*}

Due to the proximity to the Galactic Plane, there exists considerable H$\alpha$ emission from bright \ion{H}{2} regions \citep{seon11}. The brightest star - HIP 88149, B2 star - capable of ionizing the warm interstellar medium is responsible for an extended reflection nebula clearly visible in the region \citep{jyo15}. Generally, the H$\alpha$ emission traces the warm ionized gases and we expect a tight correlation between H$\alpha$ and FUV emissions if and only if the region is optically thin. Here, we notice a weak saturation of FUV intensities with H$\alpha$ emission, with a correlation coefficient, r = 0.28 (Fig.~\ref{fig8}) as compared with atomic or total (atomic + molecular) hydrogen column density (see Fig.~\ref{fig2} \& Fig.~\ref{fig5}). This is because some fraction of the H$\alpha$ intensities shows correlation with FUV emission especially in the relatively low dense regions due to the lower extinction cross-section at H$\alpha$ than at FUV wavelength which is in agreement with \citet{park12}.

\begin{figure*}
\centering
\includegraphics[width=4in]{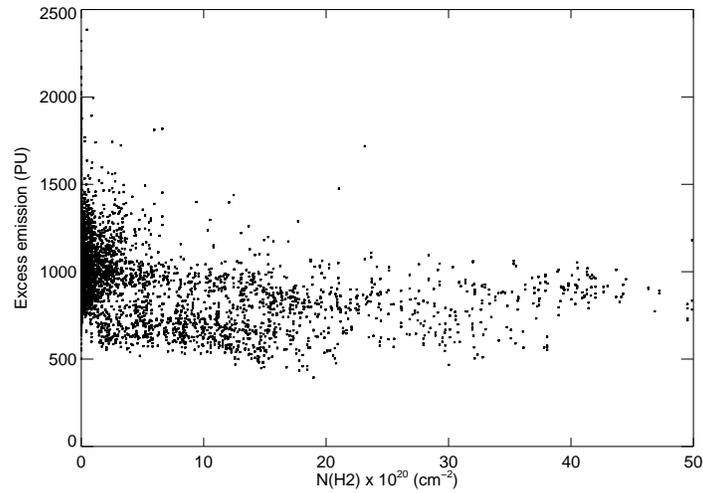}
\caption{Excess FUV emission plotted against molecular hydrogen column density, N(H$_{2}$). Poor correlation indicate the composite nature of excess FUV emission.}
\label{fig9}
\end{figure*}

\begin{figure*}
\centering
\includegraphics[width=4in]{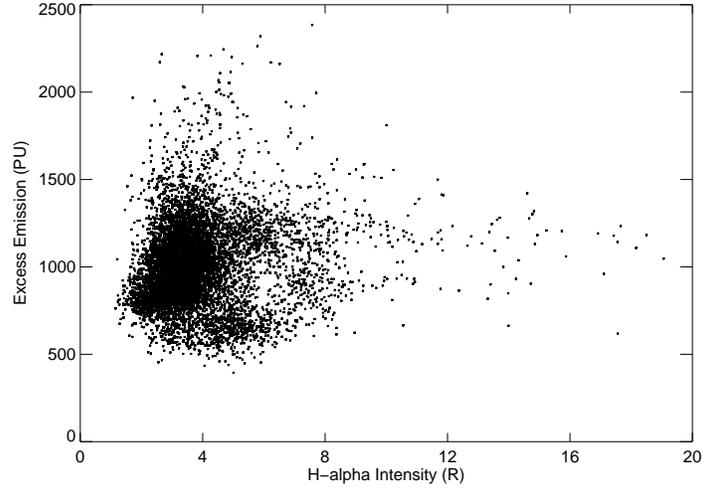}
\caption{Excess FUV emission plotted against H$\alpha$ intensity. It is clear from the figure that H$\alpha$ shows a better dependence on the excess FUV emission than N(H$_2$).}
\label{fig10}
\end{figure*}

\begin{figure*}
\centering
\includegraphics[width=4in]{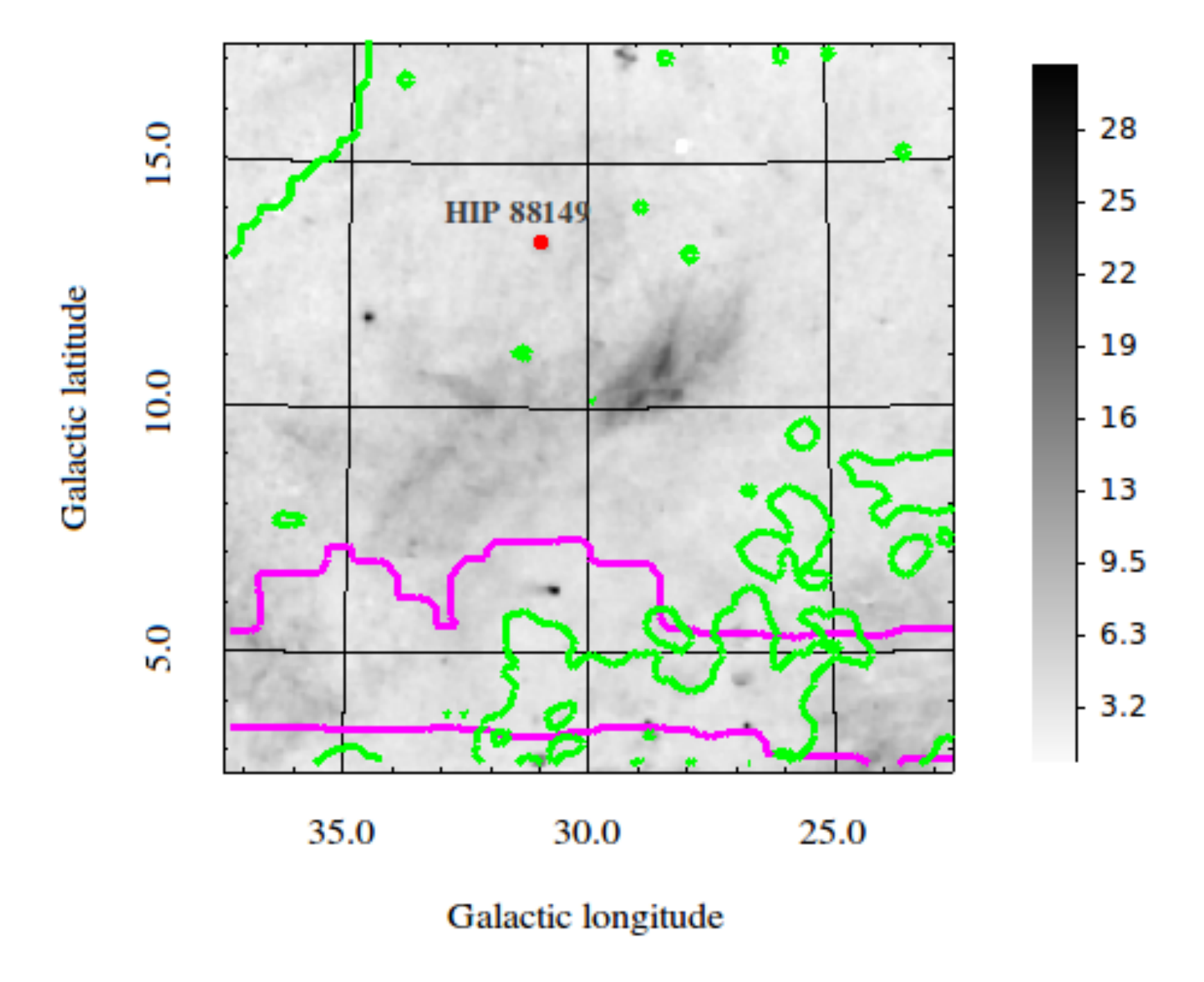}
\caption{CO (green) and N(\ion{H}{1}) (magenta) contours overlaid on the H$\alpha$ image of the region. The position of the brightest star is marked in red. It is clear from the figure that the contours corresponding to individual emission peaks stands separately.}
\label{fig11}
\end{figure*}

Fig.~\ref{fig9} shows that the excess FUV emission is highly saturated with the molecular hydrogen column density, N(H$_{2}$) in regions with optical depth, $\tau \ \ge$ 1, whereas the excess emission is weakly saturated with H$\alpha$ emission from ionized gas in the region (Fig.~\ref{fig10}). In effect we can say that the excess FUV emission estimated here is composite in nature with contributions from: (1) line emissions from hot interstellar gas, \ion{Si}{2} or \ion{C}{4} (2) two photon continuum emissions and (3) the H$_{2}$ fluorescence. 
In Fig.~\ref{fig11}, we  overlaid the CO contour (green) and N(\ion{H}{1}) contour (magenta) on an H$\alpha$ image. CO is used as the tracer of H$_2$ in such a way that N(\ion{H}{1}) is a tracer of dust scattered emission. It is clear from the figure that CO, N(\ion{H}{1}) and H$\alpha$ emission peaks are clearly distinct and the H$\alpha$ emission peaks near the reflection nebula in the region \citep{jyo15}, for which a moderate latitude dependence is also visible. The H$_{2}$ (traced by CO) emission is mainly from the lower latitudes. The H$\alpha$ emission does not depend on either N(H$_{2}$) or N(\ion{H}{1}), but only on the presence of bright stars in the region. This indicates that various components which contribute to the observed excess FUV emission in the field originate from distinct regions and hence are too complex to isolate.

Our region clearly coincides with the ‘Ophiuchus’ region and a major part of the Aquila Serpens star forming region as described by \citet{park12}, where the presence of molecular hydrogen fluorescence, two photon continuum emissions and line emission from ionized gas are reported. We suggest that the excess FUV emission from the region is composite in nature with contributions from line emissions from hot interstellar gas, \ion{Si}{2} or \ion{C}{4} in addition to HII 2photon emission and H2 fluorescence. Only a complete spectroscopic analysis of the region would be able to identify and separate the individual contributions.

\section{Conclusions}

We have analyzed the FUV (1350 - 1750 \AA)  \& NUV (1750 - 2850 \AA) emissions towards the Aquila Rift using {\it GALEX}  observations and have conducted a correlation studies using multi-wavelength data from various telescopes. The main conclusions of this study are:

\begin{enumerate}
\item The median diffuse background in the region is estimated as 2040 PU and 1718 PU in FUV and NUV respectively with an error $\le$ 40 PU. The FUV emission peaks in the relatively low-density regions which are the surrounding regions of Aquila Rift.
\item The observed FUV and NUV intensities in the region correlate well with each other. Its latitude dependence is studied and empirical equations are derived.
\item 	UV/FIR ratio in the region drops off exponentially with the IR 100 $\micron$ emission highly similar to that observed in different optically thick regions of the sky by various authors. 
\item 	The majority of the observed diffuse UV emissions are due to the dust scattered starlight which correlates well with the amount of dust in regions, where N(H) $\le$ $\sim$ 2.0 x 10$^{21}$ and is highly saturated beyond N(H) $\sim$ 2.0 x 10$^{21}$ due to the heavy dust extinction.
\item  The observed FUV intensity correlates reasonably well with H$\alpha$ emission in the faint H$\alpha$ regions and the correlation decreases as the H$\alpha$ emission increases. This may possibly indicate that the contribution of line emission from ionized gas in the FUV band is relatively high from the faint H$\alpha$ regions.
\item 	An excess emission in FUV is estimated to be in the range of 400 - 2700 PU over the region calculated by fixing an empirical FUV/NUV ratio as 0.6. This was plausibly due to contributions from fluorescent emission from the Lyman band of molecular hydrogen, line emission from ionized gas and 2-photon emission. 
\item 	The contour analysis shows that N(H$_{2}$) (traced by CO) and H$\alpha$ emissions are spread all over the field and their peak emission regions are clearly separated. This indicates that the excess FUV emission in the region is complex and composite in nature and the individual contributions can be separated only by spectroscopic analysis in FUV.
\end{enumerate}

\section{Acknowledgments}

We thank our anonymous referee for the helpful comments
and suggestions which have significantly improved this paper. This work is based on the data from NASA's {\it GALEX} spacecraft. {\it GALEX} is operated for NASA by the California Institute of Technology under NASA contract NAS5-98034. We acknowledge the NASA's Astrophysics Data System (http://cdsads.u-strasbg.fr/) and NASA's SkyView facility 
(http://skyview.gsfc.nasa.gov) located at NASA Goddard Space Flight Center. Both SNJ and NK is grateful to Sri Mata Amritanandamayi Devi, the Chancellor of Amrita Vishwa Vidyapeetham for her kind patronage and inspiration to do the  work and thanks Prof. Jayant Murthy, IIA Bangalore, for his support and guidance. SNV thanks the support of DST-FIST at SXC, Aluva.

\clearpage

\begin{deluxetable}{cc}

\tabletypesize{\scriptsize}
\tablecaption{Details of the region\label{table1}}
\tablewidth{0pt}
\tablehead{
\colhead{Parameter} & \colhead{Value}
}

\startdata
{\it l} (center) & $30\degr$ \\
{\it b} (center) & $10\degr$ \\
Area coverage   & 15 deg$^2$  \\
No:of FUV observations & 227 \\
No:of NUV observations & 258 \\
No: of FUV visits & 399 \\
No: of NUV visits & 567 \\
\enddata

\end{deluxetable}

\clearpage


\begin{thebibliography}{}

\bibitem[Dame et al.(2001)]{dame01} Dame, T.~M., Hartmann, D., \& Thaddeus, P.\ 2001, \apj, 547, 792
\bibitem[Dickey \& Lockman(1990)]{dick90} Dickey, J.~M., \& Lockman, F.~J.\ 1990, \araa, 28, 215
\bibitem[Edelstein et al.(2006)]{edel06} Edelstein, J., Min, K.-W., Han, W., et al.\ 2006, \apjl, 644, L153
\bibitem[Finkbeiner(2003)]{Fink03} Finkbeiner, D.~P.\ 2003, \apjs, 146, 407
\bibitem[Hamden et al.(2013)]{Hamden13} Hamden, E.~T., Schiminovich, D., \& Seibert, M.\ 2013, \apj, 779, 180
\bibitem[Hayakawa et al.(1969)]{hay69} Hayakawa, S., Yamashita, K., \& Yoshioka, S.  1969, \apss, 5, 493
\bibitem[Jyothy et al.(2015)]{jyo15} Jyothy, S.~N., Murthy, J., Karuppath, N., \& Sujatha, N.~V.\ 2015, \mnras, 454, 1778
\bibitem[Lillie \& Witt (1976)]{lil76} Lillie, C. F. \& {Witt}, A. N. 1976, \apj, 208, 64
\bibitem[Madsen \& Reynolds(2005)]{madsen05} Madsen, G.~J., \& Reynolds, R.~J.\ 2005, \apj, 630, 925
\bibitem[Martin et al.(2005)]{martin05} Martin, D.~C., Fanson, J., Schiminovich, D., et al.\ 2005, \apjl, 619, L1
\bibitem[Morrissey et al.(2007)]{morr07} Morrissey, P., Conrow, T., Barlow, T.~A., et al.\ 2007, \apjs, 173, 682
\bibitem[Murthy et al.(2001)]{murthy01} Murthy, J., Henry, R.~C., Paxton, L.~J., \& Price, S.~D.\ 2001, Bulletin of the Astronomical Society of India, 29, 563
\bibitem[Murthy(2014a)]{murthy14a} Murthy, J.\ 2014a, \apss, 349, 165
\bibitem[Murthy(2014b)]{murthy14b} Murthy, J.\ 2014b, \apjs, 213, 32
\bibitem[Park et al.(2012)]{park12} Park S.-J. Min K.-W. Seon K.-I. Han W. Lee D.-H. Edelstein J., et al.\ 2012, \apj, 75, 10
\bibitem[Perault et al.(1991)]{per91} Perault, M., Lequeux, J., Hanus, M., \& Joubert, M.\ 1991, \aap, 246, 243 
\bibitem[Sasseen et al.(1995)]{sas95} Sasseen, T.~P., Lampton, M., Bowyer, S., \& Wu, X.\ 1995, \apj, 447, 630 
\bibitem[Schiminovich et al.(2001)]{schim01} Schiminovich, D., Friedman, P.~G., Martin, C., \& Morrissey, P.~F.\ 2001, \apjl, 563, L161 
\bibitem[Schlegel et al.(1998)]{schlegel98} Schlegel, D.~J., Finkbeiner, D.~P., \& Davis, M.\ 1998, \apj, 500, 525  
\bibitem[Seon et al.(2011)]{seon11} Seon, K.-I., Witt, A., Kim, I.-J., et al.\ 2011, \apj, 743, 188 
\bibitem[Shalima et al.(2006)]{sha06} Shalima, P., Sujatha, N.~V., Murthy, J., Henry, R.~C., Sahnow, D.~J.\ 2006, \mnras, 367 (4), 1686
\bibitem[Strai{\v z}ys et al.(2003)]{str03} Strai{\v z}ys, V., {\v C}ernis, K., \& Barta{\v s}i{\= u}t{\.e}, S.\ 2003, \aap, 405, 585
\bibitem[Sujatha et al.(2010)]{suj10} Sujatha, N.~V., Murthy, J., Suresh, R., Conn Henry, R., \& Bianchi, L.\ 2010, \apj, 723, 1549 


\end{thebibliography}
\end{document}